\documentstyle[sprocl,epsbox]{article}

\def\Journal#1#2#3#4{{#1} {\bf #2}, #3 (#4)}

\def\PLB{{\em Phys. Lett.}  B}

\def\slash#1{{\mathpalette\c@ncel{#1}}} 
\def\pslash{\rlap/{\mkern-1mu p}}
\def\nslash{\rlap/{\mkern-1mu n}}

\def\be{\begin{equation}}
\def\ee{\end{equation}}
\def\bea{\begin{eqnarray}}
\def\eea{\end{eqnarray}}

\begin{document}

\title{SINGLE-SPIN ASYMMETRIES AND SOFT-GLUON POLES
\footnote{Talk presented at the conference DIS2001 held at Bologna, Italy,
from April 27 to May 1, 2001.}}

\author{YUJI KOIKE}

\address{Department of Physics, Niigata University, Niigata 950-2181,
Japan\\E-mail: koike@nt.sc.niigata-u.ac.jp} 


\maketitle\abstracts{Mechanism for the
single transverse spin asymmeties in the pion production,
$p^\uparrow + p'\to \pi(\ell_T) + X$, 
and the $\Lambda$ hyperon polarization, 
$p + p' \to \Lambda^\uparrow (\ell_T)+ X$, is investigated in 
the framework of QCD factorization theorem.
We identify all the soft-gluon pole contributions
coming from the twist-3
distribution and/or fragmentation functions
and show that they give rise to large asymmetries at large $x_F$.
}

\section{Introduction}
The interest in the
single transverse spin asymmety $A_N$ in the pion production,
$N(P,S_\perp) + N(P')\to \pi(\ell_T) + X$,
and the hyperon (typically 
$\Lambda$) polarization $P_\Lambda$ in the unpolarized
$NN$ collision,
$N(P) + N(P') \to \Lambda(\ell_T,S_\perp) + X$, 
resides in the fact that
they probe quark-gluon correlation in the hadrons
(higher twist effect) 
which is not included in the parton model. 
Without it, the partonic cross sections are strongly
suppressed by $m_q/Q$ ($m_q$ is a quark mass and $Q$ is a hard scale involved)
and gives negligible asymmetries.

In this talk,  I will discuss $A_N$ and $P_\Lambda$
in the framework of the collinear factorization.  According to
the generalized QCD factorization theorem,
the cross section for $A_N$ typically 
consists of three kinds of twist-3 cross sections, 
\bea
&{\rm (A)}&\quad G_a(x_1,x_2)\otimes q_b(x')\otimes 
\widehat{q}_{c\to \pi}(z)\otimes
\hat{\sigma}^1_{ab\to c},
\label{eq1}\\
&{\rm (B)}&\quad
\delta q_a(x)\otimes E_b(x_1',x_2') \otimes 
\widehat{q}_{c\to \pi}(z)\otimes 
\hat{\sigma}^2_{ab\to c},\\
&{\rm (C)}&\quad
\delta q_a(x)\otimes q_b(x') \otimes 
\widehat{E}_{c\to \pi}(z_1,z_2)\otimes
\hat{\sigma}^3_{ab\to c},
\eea  
and $P_\Lambda$ likewise receives two contributions,
\bea
&{\rm (A')}&\quad E_a(x_1,x_2)\otimes q_b(x')\otimes 
\delta \widehat{q}_{c\to \Lambda}(z)\otimes
\hat{\sigma}^4_{ab\to c},\\
&{\rm (C')}&\quad
q_a(x)\otimes q_b(x') \otimes 
\widehat{G}_{c\to \Lambda}(z_1,z_2)\otimes
\hat{\sigma}^5_{ab\to c}.
\label{eq2}
\eea  
Here the functions with two variables (momentum fractions)
$G_a(x_1,x_2)$, $E_a(x_1,x_2)$, $\widehat{E}_{c\to \pi}(z_1,z_2)$
$\widehat{G}_{c\to \Lambda}(z_1,z_2)$ are twist-3 quantities:
$G_a$ and $E_a$ are, respectively,  
the transversely polarized distribution
and the unpolarized distribution functions in the nucleon.
The functions with a hat, 
$\widehat{E}_{c\to \pi}$ and 
$\widehat{G}_{c\to \Lambda}$ are, respectively,
the unpolarized fragmentation 
function for the pion and the transversely polarized 
fragmentation function for $\Lambda$.
$a$, $b$ and $c$ stand for the parton's species.
Other functions are twist-2; 
$q_b(x)$ the unpolarized distribution (quark or gluon), 
$\delta q_a(x)$ the transversity distribution,
$\widehat{q}_{c\to\pi}$ the unpolarized fragmentation function and 
$\delta\widehat{q}_{c\to\Lambda}$ the transversity fragmentation
function for $\Lambda$.  
$\hat{\sigma}^1_{ab\to c}$ {\it etc} represents the partonic cross section
for the process
$a+b \to c + anything$ which yields large transverse momentum of
the parton $c$. 
Note that 
$\delta q_a$, $E_a$, $\delta\widehat{q}_{c\to\Lambda}$ and
$\widehat{E}_{c\to\pi}$
are chiral-odd, and (B), (C) and (A') contain two chiral-odd functions,
which should appear in a pair along a fermion line in the 
diagram for the cross sections.

\section{Twist-3 distribution and fragmentation functions}
Relevant
twist-3 distributions are defined from a 
quark-gluon correlation in the nucleon
\bea
M_{Fij}^\alpha (x_1,x_2)
=\int {d\lambda\over 2\pi}\int {d\mu\over 2\pi} e^{i\lambda x_1}
e^{i\mu (x_2-x_1)} \langle PS|\bar{\psi}_j(0)
gF^{\alpha +}(\mu n)\psi_i(\lambda n)
|PS\rangle,
\eea
and they are classified in eqs.(7)-(9) of Ref.\cite{KKp1}.
Likewise the twist-3 fragmentation functions are defined from
\bea
\widehat{M}_{Fij}^\alpha (z_1,z_2) &=&\sum_X
\int {d\lambda\over 2\pi}\int {d\mu\over 2\pi} e^{-i\lambda/z_1}
e^{-i\mu (1/z_2-1/z_1)}\nonumber\\
&\times&\langle 0 |\psi_i(0)|HX\rangle
\langle HX | gF^{\alpha \beta}(\mu n) n_\beta\bar{\psi}_j(\lambda n) | 0\rangle.
\label{frag}
\eea
Similarly to ${M}_F^\alpha (x,y)$,
$\widehat{M}_F^\alpha (z_1,z_2)$ is decomposed
to define twist-3 fragmentation funcitons as
\bea
\widehat{M}_F^\alpha (z_1,z_2)
&=& M/4 \pslash \epsilon^{\alpha p n S_\perp} \widehat{G}_F(z_1,z_2)/z_2
-iM/4\pslash \gamma_5 S_\perp^\alpha \widehat{G}^5_F(z_1,z_2)/z_2
\nonumber\\
&-&iM(S\cdot n)/4 \gamma_5 (p^\alpha\nslash\pslash -\gamma^\alpha\pslash)
\widehat{H}_F(z_1,z_2)/z_2\nonumber\\
&+&M/4 
\gamma_5 \pslash\gamma_\nu \epsilon^{\nu\alpha n p}\widehat{E}_F(z_1,z_2)
/z_2+\cdots,
\eea
where $\cdots$ stands for the twist higher than 3.
In (\ref{frag}), if one shifts the gluon field strength 
$gF^{\alpha \beta}(\mu n) n_\beta$ into the matrix element
with $\psi(0)$ and call it $\widehat{M}^\alpha_{FR}(z_1,z_2)$, 
similar decomposition of $\widehat{M}^\alpha_{FR}$
defines another fragmentation 
functions 
($\widehat{G}_{FR}$, $\widehat{G}^5_{FR}$, $\widehat{H}_{FR}$,
$\widehat{E}_{FR}$).  But these functions are related to 
($\widehat{G}_{F}$, $\widehat{G}^5_{F}$, $\widehat{H}_{F}$,
$\widehat{E}_{F}$) by the hermiticity.  In addition if we assume naive time 
reversal invariance, 
these functions become real and obey the relation 
$\widehat{G}_{FR}(z_1,z_2) = \widehat{G}_F(z_2,z_1)$,
$\widehat{E}_{FR}(z_1,z_2) = \widehat{E}_F(z_2,z_1)$, 
$\widehat{G}^5_{FR}(z_1,z_2) = -\widehat{G}^5_F(z_2,z_1)$,
$\widehat{H}_{FR}(z_1,z_2) = -\widehat{H}_F(z_2,z_1)$.
We assume this symmetry property in our analysis.

\section{Result for the asymmeties}

With the complete set of the distribution and fragmentation functions up to
twist-3, one can derive the cross section formula corresponding to
(\ref{eq1})-(\ref{eq2}).  We follow the previous analyses
Ref.\cite{QS91} - Ref.\cite{KKh} and 
employ the {\it valence-quark soft-gluon} approximation
to analyze the cross section.
In this approximation we keep only the terms with the derivative
of the twist-3 functions
such as $d E_F(x,x)/dx$ and $d\widehat{E}_F(z,z)/dz$.
This approximation should be valid 
at large $x_F\to 1$, which probe 
the region with large $x$, small $x'$
and large $z$ in (\ref{eq1})-(\ref{eq2}) where the relations such as  
$|dE_F(x,x)/dx| >> E_F(x,x)$ and 
$|d\widehat{E}_F(z,z)/dz| >> \widehat{E}_F(z,z)$ hold.  
The (B) term for $A_N$ may cause enhancement in the asymmetry at
$x_F \to -1$, but it turns out that it is negligible in all kinematic region
because of the smallness of the hard cross section 
$\hat{\sigma}^2$\,(Ref.\cite{KKp2}).

To get a rough feeling on how each term for $A_N$ and $P_\Lambda$
behaves, we show the (C) term for $A_N$ in Fig.1, and (A') and (C')
term for $P_\Lambda$ in Fig. 2.  We use the same distribution and 
fragmentation functions used in Refs.\cite{KKp2,KKh}.  For the twist-3
functions, we make an extention of the ansatz 
taken in Ref.\cite{QS99}-Ref.\cite{KKh},
$G_F^a(x,x)=K_a q^a(x)$,
$E_F^a(x,x)=K'_a \delta q^a(x)$,
$\widehat{G}_F^a(x,x)=\widetilde{K}_a \widehat{q}^a(x)$,
$\widehat{E}_F^a(x,x)=\widetilde{K}'_a \widehat{q}^a(x)$,
and set $K_u=-K_d=0.07$, $K'_a=\widetilde{K}_a = K_a$ ($a=u,d$),
and $\widetilde{K}'_u=-0.11$, $\widetilde{K}'_d=-0.19$.
This choice of $\widetilde{K}'_a$ is simply motivated to 
reproduce $A_N$ approximately at large $x_F$.  

\begin{figure}[h]
\center
\epsfile{file=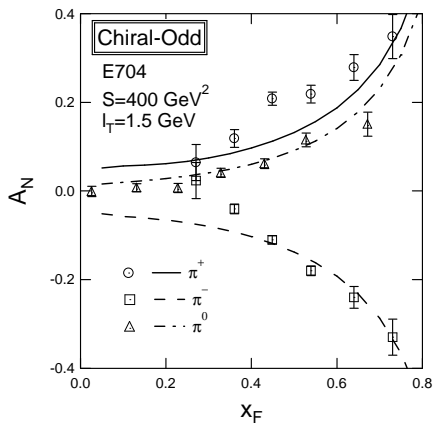,scale=0.45}
\caption{(C) contribution to $A_N$ for $\pi^{\pm,0}$.  \label{figB}}
\end{figure}

\begin{figure}[h]
\center
\epsfile{file=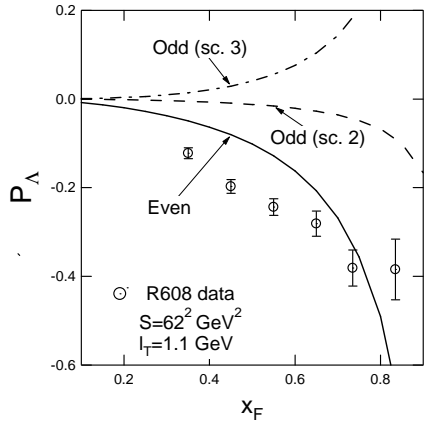,scale=0.45}
\caption{
(A') (Chiral-odd with Scenarios 2 and 3 for $\delta\hat{q}$,
and (C') (Chiral-even) contributions to $P_\Lambda$.}
\end{figure}

One sees from Fig.1 that the (C) term alone can give equally good fit
to the E704 data as the (A) term studied in Ref.\cite{QS99}.
The (C') contribution also gives rise the rising behavior 
of $P_\Lambda$ at large $x_F$.

To summerize, we have presented an analysis for $A_N$ and $P_\Lambda$
in the framework of QCD factorization, in particular, the soft gluon-pole 
contribution with the twist-3 fragmentation functions is identified.
With a moderate model assumption
for the twist-3 fragmentation fuctions, 
derivative of the twist-3 fragmentation function also gives
the rising behavior of the asymmetry at large $x_F$.

\section*{Acknowledgments}
I would like to thank George Sterman for helpful discussions.
This work is supported in part by RIKEN and the Grant-in-Aid for Scientific 
Research (No. 12640260) of Monbusho.

\end{document}